\documentclass[11pt,leqno]{article}%
\usepackage{amssymb,bbm,amsmath,
amsthm,
amscd}
\usepackage{vmargin}
\usepackage[dvipsnames]{xcolor}
\usepackage{
 array}
 
 \usepackage{graphicx}
\catcode`\@=11
\newbox\bz@
\newdimen\bdimz@
\def\linethrough#1{\setbox\bz@=\hbox{#1}%
\bdimz@=\ht\bz@ \divide\bdimz@ by 5 \advance\bdimz@ by -\dp\bz@ \ht\bz@=\bdimz@
\leavevmode\hbox{$\overline{\overline{\box\bz@}}$\relax}}
\catcode`\@=12

\def \new #1 {\textcolor{Maroon}{\underline{#1}} }

\usepackage[colorlinks=true]{hyperref}
\newcommand{\BEQ} {\begin{equation} }
  \newcommand{\EEQ} {\end{equation} }
\newcommand{\fin}{\hfill $\Box$}
\newcommand{\BTHM}{\begin{theorem}}
  \newcommand{\ETHM}{\end{theorem}}
\hypersetup{urlcolor=blue, citecolor=red}

\newtheorem{lemma}{Lemma}[section]
\newtheorem{corollary}{Corollary}[section]

\newtheorem{theorem}{Theorem}[section]
\newtheorem{definition}{Definition}[section]

\newtheorem{remark}{Remark}[section]

\newtheorem{Corollary}{Corollary}[section]

\newcommand{\tore}{\mathbb{T}_3}

\newcommand{ \E}{\varepsilon}
\newcommand{ \vit}{\hbox{\bf u}}

\newcommand{ \vittest }{\hbox{\bf v}}
\newcommand{ \wit}{\hbox{\bf w}}

\newcommand{ \bu}{\hbox{\bf u}}
\newcommand{ \bx}{\hbox{\bf x}}

\newcommand{ \bw}{\hbox{\bf w}}

\newcommand{ \bff}{\hbox{\bf f}}
\newcommand{ \Z}{\mathbbm{Z}}
\newcommand{\moy} {\overline {\vit} }
\newcommand{\g} {\nabla }
\newcommand{\p} {\partial}
\newcommand{\x} {{\bf x}}
\newcommand{\N}{\mathbb{N}}
\newcommand{\R}{\mathbb{R}}

\newcommand{\err}{\boldsymbol{\varepsilon}}
\newcommand{\res}{\boldsymbol{\tau}}
\newcommand{\woy}{\overline \wit} 
\newcommand{\pre}{\overline p}

\newcommand{\esp}{\mathbb{H} }

\makeatletter 

\newcounter{taskcounter}[section]

\newcommand{\step}{ \refstepcounter{taskcounter} { 
{\bf Step \arabic{section}.\roman{taskcounter}}.  }
}

\@addtoreset{equation}{section}

\begin{document}

\title{Modeling error in Approximate Deconvolution Models}
\author{Argus A. Dunca\thanks{Faculty of Mathematics and Computer Science, Spiru Haret University, Bucharest, 030045, Romania,
a.a.dunca.mi@spiruharet.ro}  \and Roger
  Lewandowski\thanks{IRMAR, UMR 6625, Universit\'e Rennes 1, Campus
    Beaulieu, 35042 Rennes cedex FRANCE;
    Roger.Lewandowski@univ-rennes1.fr,
    http://perso.univ-rennes1.fr/roger.lewandowski/}} \date{}
\maketitle

\begin{abstract} 
We investigate the assymptotic behaviour of the
modeling error in approximate deconvolution model in the 3D periodic case, when the order $N$ of deconvolution goes to $\infty$. We consider successively the  generalised Helmholz filters of order $p$ and the Gaussian filter. For Helmholz filters, we estimate the rate of convergence to zero thanks to energy budgets, Gronwall's Lemma and sharp inequalities about Fouriers coefficients of the residual stress. We next show why the same analysis does not allow to conclude convergence to zero of 
the error modeling in the case of Gaussian filter, leaving open issues. 

 \end{abstract}

MCS Classification : 76D05, 35Q30, 76F65, 76D03

\medskip

\section{Introduction}

Direct Numerical Simulations of flows  from 
the Navier-Stokes Equations (NSE) 
\begin{equation}\label{nse}
\begin {array}{rcl}
   \bu_{t}+\nabla \cdot (\bu\otimes \bu)  - \nu \Delta \vit+\nabla p &= &\bff, \\
   \nabla\cdot \bu&= & 0,\\
   \bu({\bf x}, 0)&= & \bu_{0}({\bf x}),
 \end{array}
\end{equation}
are accurate only for small Reynold numbers. For large Reynolds numbers, flows are turbulent and only means or large scales 
of  velocity and pressure fields might be computed thanks to turbulent models.
\medskip 

Large Eddy Simulation (LES) modeling of turbulent flows
aims to apply to the NSE a low pass filter specified by a convolution kernel $G$, leading to the filtered NSE, written in the form 
\begin{equation}\label{smago} \begin {array}{rcl}
\overline{\bu}_{t}+\nabla \cdot (\overline{\bu}\otimes \overline{\bu})   
- \nu \Delta \moy
+\nabla \overline{p} & = & \overline{\bff} +\nabla \cdot \mathbb{S} (\bu,\bu), \\
   \nabla\cdot \moy&= & 0,\\
   \moy({\bf x}, 0)&= & \overline{ \bu_{0}}({\bf x}), 
   \end{array}
\end{equation}
where $ \moy = G \star \vit$ is the large scale velocity, $\overline p = G \star p$ the large scale pressure,  
\BEQ \mathbb{S} (\bu,\bu)= \overline{\bu}\otimes \overline{\bu}-\overline{\bu\otimes \bu}, \EEQ
is the subfilter scale stress tensor. A modelisation process aims to seek for suitable approximations 
to $\mathbb{S} (\bu,\bu)$ in terms of $\moy$ to close System (\ref{smago}), that yields a LES model  \cite{BIL2006, MG00, Sag2001}.

\medskip

 Most of LES models are over diffusive and trend to underestimate the energy, creating a subfilter scale  region (SFS). 
 The total error committed   is the sum of  the numerical error NE and the SFS area  \cite{CSXF05}. 
To reduce the SFS area, one uses to apply a deconvolution  operator to the filter \cite{CSXF05, GC03, SAK2001, LL2006b, LR2012}.  

\medskip

 \begin{figure} \label{fig:SGS_SFS} \begin{center} {\includegraphics
 [scale=0.6]{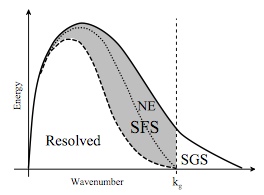}}  \label{fig:scales}\caption{\it From Chow
 et al. 2005 \cite{CSXF05}. American Meteorological Society. Reprinted with permission. }\end{center}   \end{figure}

The aim of this paper is
to estimate the error  modeling in terms of the order of the deconvolution denoted by $N$, in the case of the simplified Bardina's model
 \cite{BFR80, LL2006a,CLT2006}, which is based on the approximation 
 \BEQ \label{eq:app_dec_0} \mathbb{S} (\bu,\bu) \approx  \mathbb{S}(\moy, \moy)= \moy \otimes \moy - \overline {\moy \otimes \moy}. \EEQ
The approximate deconvolution model (ADM in what follows) is deduced from the simplified Bardina's model by changing approximation (\ref{eq:app_dec_0}) in 
 \BEQ \label{eq:app_dec} \mathbb{S} (\bu,\bu) \approx    \mathbb{S}_N (\moy, \moy) = \moy \otimes \moy - \overline {D_N(\moy) \otimes D_N (\moy)}, \EEQ
 where the deconvolution operator $D_N$ is such that 
  \BEQ D_N = \sum_{n = 0}^N (I-G)^n, \EEQ
  while still noting $G$ the operator associated to the kernel $G$. 
We always have $ \mathbb{S}_0 (\moy, \moy) =  \mathbb{S} (\moy, \moy)$, and when $ || G || <1$\footnote{the operator norm is based on natural energy spaces the fields belongs to, which will be specified latter} then for a fixed
$\vit$,
\BEQ \label{eq:sfs_conv} \lim_{N \rightarrow \infty }Ê \mathbb{S}_N (\moy, \moy)  = \mathbb{S} (\bu,\bu). \EEQ
Let  $(\moy_N, \pre_N )$ be the field calculated from approximation (\ref{eq:app_dec}), that is  the solution to the system
  \begin{equation} \label{eq:adm11}
  \begin{aligned}
   \partial_t \moy_N +\nabla\cdot(\overline {D_{N} (\moy_N) \otimes
     D_{N} (\moy_N)})-\nu\Delta \moy_N +\nabla \pre_N = \overline \bff,   
   \\
   \nabla\cdot \moy_N =0,   
   \\
 \moy_N (0, \x) = \overline{\vit_0} (\x),
 \end{aligned}
\end{equation}
if any solution exists. 
Existence and uniqueness of a solution to System (\ref{eq:adm11}) was first proved in \cite{DE2006} when $G$ is the usual Helmholz filter in the 3D periodic case.  More generally, if 
one can prove existence and uniqueness of a solution to system  (\ref{eq:adm11}) for any $G$ that satisfies
(\ref{eq:sfs_conv}), it is expected that the sequence $(\moy_N, \pre_N)_{N \in \N}$ converges to 
$(\moy, \pre) = (G \vit, G p)$, for some solution $(\vit, p)$ of the NSE. \medskip

Such convergence results  has been proved 
in \cite{BL11}  in the 3D periodic case, when $G= G_{\alpha,p}$ is the generalised Helmholz filter of order $p$ with $p \ge 3/4$, where
\begin{equation} \label{eq:transfer_p} G_{\alpha,p}(\x)  =\sum_{\mathbf{k} \in {\cal T}_3^\star}
{e^{i\mathbf{k \cdot x}} \over 1 + \alpha ^{2p} |{\bf k}|^{2p}},  \footnote{In terms of operators 
$G_{\alpha, p}  = ({\rm I} - \alpha^{2p} \Delta^p) ^{-1}$, where $\Delta^p$ denotes the $p$-Laplacien, and 
$A_{\alpha,p} = {\rm I} - \alpha^{2p} \Delta^p= G^{-1}$}
\end{equation}
after having proved existence and uniqueness of $(\moy_N, \pre_N)$. In Definition (\ref{eq:transfer_p}),  ${\cal T}_3 := 2 \pi \Z^3 /L$, $L >0$ being the size of the computational box, and $\alpha >0$ is the filter's width, usually of same magnitude of the mesh size in a numerical simulation  (see \cite{KL92} for further discussions). 
\medskip

This yields to consider
the error modeling 
$\boldsymbol{\E} _N = \moy - \moy_N, $
which goes  to zero when $N$ goes to infinity.  It remains the issue of estimating the rate of convergence in terms of $N$. 
Staying within the 3D periodic  framework and   the generalised Helmholz filter of order $p$ ($p \ge 3/4$), we show in  this paper that  $L^2$ and $H^1$ norms of 
$\err_N$ are of order $ (p (N+1))^{-1/4p}$, (see our main result, Theorem \ref{thm:main} below).  
\medskip

To derive this rate of convergence, we first write the equation satisfied by $\boldsymbol{\E} _N$, by substracting  (\ref{eq:adm11}) to (\ref{smago}), which yields
\BEQ \label{eq:equation_for_err1}  \p_t \err_N + \g \cdot (\overline{ D_N \err_N \otimes D_N \moy_N }) -\nu \Delta \err_N + \g r_N = - \g \cdot \overline \res_N - 
\g \cdot \overline{D_N \moy \otimes D_N \err_N},
\EEQ
where $r_N = \pre -\pre_N$, and 
\BEQ \label{eq:res_stress} \res_N = \vit \otimes \vit - D_N \moy\otimes D_N \moy \EEQ
is the residual stress. By using successively an energy budget procedure and Gronwall's Lemma, we get  an inequality satisfied by the norms of 
$A^{1/2}D_N \err_N$ where $A = G^{-1}$ (in terms of operators), from which we deduce an inequality satisfied by the norms of 
$\err_N$ itself (see  Inequality (\ref{eq:cor_energy}) below).  This inequality  highlights the role played by the $L^2$ norm of the residual stress. 
\medskip

The weakness of this method 
is the regularity assumption that should be imposed on the field $\vit$, which should be in $L^4 ( H^1)$. However, such proceedings are similar to 
usual uniqueness proofs about the NSE, always involving regularity assumptions. 

\medskip

It remains to estimate the $L^2$ norm of the residual stress (see Inequality (\ref{eq:est_res2})). We carry out this calculation by using Fourier series expansion and  calculations outlined in Appendix \ref{sec:appendix}, which, if they use elementary real analysis only, are not straightforward and were first speculated thanks to numerical and symbolic computations, before being rigorously proved. 

\medskip

We observe that the rate of convergence slows down as $p$ increases in the range $[1, \infty[$. Moreover, the resulting bound 
 goes to a constant that 
only depends on $\alpha$ and $\vit$ when $p$ goes to infinity and $N$ remains fixed. This is consistent with the idea that more large is
$p$, then more smooth are the filtered fields, which should enlarge the SFS area.  Therefore, one needs high orders of deconvolution to reconstruct well the 
resolved scale area for large values of $p$. 
\medskip

Then we  consider the popular Gaussian filter, 
\BEQ
\tilde G_{\alpha}(\bx) = \left(\frac{6}{\alpha^2\pi}\right)^{3/2} \exp
\left(-\frac{6}{\alpha^2}|| \x||^2\right), 
\EEQ  
often used in LES.  Applying the ADM theory for general abstract filters developed  in
\cite{stanculescu}, we deduce that the ADM is well-posed in the case of the Gaussian filter. Therefore, one may ask if there is convergence of the model to the filtered NSE when $N \rightarrow \infty$, and if yes 
what is the convergence rate. 
\medskip

The theory we develop for Helmholz filters,  does not apply to the Gaussian filter, because of a too strong convergence of its Fourier modes to zero as the wave number increases, although this is not an evidence that the convergence does not hold. 
\medskip

We argue by approximation in showing that the Gaussian filter can be approximated by 
\begin{equation} \label{eq:gauss_approx} \displaystyle \tilde G_{\alpha,m}(\x)  =\sum_{\mathbf{k} \in {\cal T}_3}
\left (1 + {\alpha ^{2} |{\bf k}|^{2} \over 24 m} \right )^{-m} e^{i\mathbf{k \cdot x}},  \footnote{In terms of operators 
$\tilde G_{\alpha,m} = \left (1 - {\alpha ^{2}  \over 24 m } \Delta \right )^{-m}$}
\end{equation}
when $m$ goes to infinity. We show that our procedure is still valid for this sequence of filters, and  we derive a bound of order 
$ (N+1) ^{-4m} $ fro them.  This bound goes to a constant depending on $\alpha$ and $\vit$ when $m$ goes to infinity for a fixed $N$. Therefore, we cannot 
conclude that the deconvolution process converges to the filtered field $(\moy, \pre)$ in the case of the Gaussian filter. Because of the strong 
regularisation effect of this filter, we may conjecture that if such a convergence would hold, then it should be very low. Therefore, the deconvolution process seems to be not appropriate for the Gaussian filter.  This remains an open issue. 
\medskip

The paper is organised as follows. We first fix the mathematical framework and recall the results of  \cite{BL11} useful for the continuation of the paper.  
We next detail how to bound the error modeling in terms of the residual stress, whose $L^2$ norm is then estimated by Fourier series expansions. 
We finally consider the Gaussian Filter by showing how to approximate it by the $G_{\alpha, m}$'s, the error modeling of which being then estimated. The paper finishes by a technical appendix including key results to derive estimates about the residual stress. 


\section{Mathematical framework}

\subsection{Space function} \label{sec:space_functions} 

Throughout the paper, $\nu>0$ and $\alpha>0$ are fixed and we stay within the periodic case framework. The domain of study  is the 3D torus 
\BEQ \mathbb{T}_3=
\R^3 / {\cal T}_3 \quad \hbox {where} \quad {\cal T}_3 := 2 \pi
\Z^3 /L, \EEQ
for some given $L>0$, which is the size of the computational box. All the fields we consider have zero mean on $\tore$. Let   $\mathbb{H}_s$ be the vector field  space 
\BEQ \label{eq:espace_de_base} 
  \mathbb{H}_s = \left \{ \wit  = (w_1, w_2, w_3) = \sum_{\mathbf{k} \in 
  {\cal T}_3^\star} \widehat \wit_{\bf k}e^{i\mathbf{k \cdot x}}:\ \  
    \sum_{\mathbf{k} \in {\cal T}_3^\star} | \mathbf{k} |^{2s} |     \widehat  {\wit }_{\bf k}|^{2} < \infty  \right \},
\EEQ 
equipped with the Hermitian structure defined by the inner product and its associated norm
\begin{equation}\label{TQW887} 
  (\wit, \hbox{\bf v} )_{s} = \sum_{\mathbf{k} \in {\cal T}_3^\star} 
  | \mathbf{k} |^{2s}  \widehat  {\wit }_{\bf k}\cdot \widehat
    {\hbox{\bf v} }_{\bf k}^\star , \quad || \wit ||_s = \left ( \sum_{\mathbf{k} \in {\cal T}_3^\star} 
  | \mathbf{k} |^{2s} |   {\wit }_{\bf k} |^2 \right )^{1 \over 2} , 
\end{equation}
where 
$$\forall \, {\bf k} = (k_1, k_2, k_3) \in {\cal T}_3, \quad | {\bf k} |^2 = k_1^2 + k_2^2 + k_3^2,$$ 
and $z^\star$ denotes the complex conjugate of $z$.  It can be proved  (see \cite{RLbookNSE}) that forall $s \in \R$, 
\BEQ  \mathbb{H}_s \hbox{ is isomorphic to } H^s(\tore)^3, \quad (\mathbb{H}_s)'  =  \mathbb{H}_{-s}, \EEQ
and we denote 
\BEQ \forall (\wit, {\bf v} ) \in \mathbb{H}_{-s} \times \mathbb{H}_{s}, \quad _{-s}(\wit, {\bf v} )_s  =  \sum_{\mathbf{k} \in {\cal T}_3^\star}  \widehat  {\wit }_{\bf k}\cdot \widehat
    {\hbox{\bf v} }_{\bf k}^\star \EEQ
    the duality pairing. 
\medskip

 Let $\mathbf{H}_{s} \subset   \mathbb{H}_s$ be the closed subspace of fields valued in $\R^3$, characterized by 
$$\mathbf{H}_{s} = \left \{  \wit  = \sum_{\mathbf{k} \in {\cal T}_3^\star} \widehat \wit_{\bf k}e^{i\mathbf{k \cdot x}} \in  \mathbb{H}_s:  \forall  \, {\bf k}Ê\in  {\cal T}_3^\star, \quad\widehat  {\wit }_{\bf k}^\star = \widehat  {\wit }_{-{\bf k}}  \, \hbox{ and } \,  
        \mathbf{k}\cdot \widehat{\bw}_\mathbf{k}=0 \right \} .$$ 
On can show (see \cite{RLbookNSE}) that 
\BEQ \label{eq:space}  \mathbf{H}_{s} = \left\{\wit : \tore \rightarrow \R^3, \, \,
   \mathbf{w} \in H^{s}(\mathbb{T}_3)^3,   \quad\nabla\cdot\mathbf{w}
   = 0, \quad\int_{\mathbb{T}_3}\mathbf{w}\,d\x = \mathbf{0} \right\},
\EEQ

\subsection{Operators}

\subsubsection{Kernel and filter} 

The general Helmholz filter $\woy = G_{\alpha, p} \star \wit $ is defined by the Fourier Series expansion of the kernel $G_{\alpha,p}$
\begin{equation} \label{eq:transfer_p1} G_{\alpha,p} (\x) =\sum_{\mathbf{k} \in {\cal T}_3^\star}
\widehat{G}_{\bf k}\,e^{i\mathbf{k \cdot x}}, \quad \widehat G_{{\bf k}} = {1 \over 1 + \alpha ^{2p} |{\bf k}|^{2p}}. 
\end{equation}
Viewed as an operator, one has $G_{\alpha,p} = ({\rm I}Ê- \alpha^{2p}Ê\Delta^{2p})^{-1}$. Furthermore, 
a given free divergence field $\wit$ being given, $\woy$ is solution  of the PDE problem
\BEQ \label{eq:gen_hemholz} \begin{array}{rcll}- \alpha^{2p} \Delta^p \overline \wit + \overline \wit + \g r&= &\wit & \hbox{in } \tore, \\
\g \cdot \wit &= & 0 & \hbox{in } \tore, \end{array} \EEQ
where the Lagrange multiplier $r$ is constant in this case.
\medskip

From now, we write $G$ instead of $G_{\alpha, p}$, and we denote in the same way kernel and operator. For all $s \ge 0$, $G$ defines an isomorphism,
\BEQ \label{eq:operator_G} G : \left \{  \begin{array} {rcl}   \mathbb{H}_s & \longrightarrow &   \mathbb{H}_{s + 2p} \\ \displaystyle
\wit  = \sum_{\mathbf{k} \in {\cal
        T}_3^\star} \widehat \wit_{\bf k}e^{i\mathbf{k \cdot x}} & \displaystyle \longrightarrow & \displaystyle  \woy  = 
        \sum_{\mathbf{k} \in {\cal T}_3^\star} \widehat{G}_{\bf k} \widehat \wit_{\bf k}e^{i\mathbf{k \cdot x}} \phantom{\int_0^1},
        \end{array} \right. \EEQ
and we set $A = G^{-1}$, characterised by its kernel
\begin{equation} \label{eq:transfer_inverse} A(\x) =\sum_{\mathbf{k} \in {\cal T}_3^\star}
\widehat{A}_{\bf k}\,e^{i\mathbf{k \cdot x}}, \quad \widehat A_{{\bf k}} = {1 + \alpha ^{2p} |{\bf k}|^{2p}}. 
\end{equation}
Notice that if $\wit \in {\bf H}_s$, then $\woy \in {\bf H}_{s+2p}$ and the restriction og $G$ to ${\bf H}_s$, still denoted by $G$ is an isomorphism 
that maps ${\bf H}_s$ onto ${\bf H}_{s+2p}$.

\subsubsection{Deconvolution} 

Let $D_N$ denote the deconvolution operator, characterised by the Kernel 
 $$\displaystyle D_N = \sum_{0 \le n \le N} ({\bf I} - G)^n=  \sum_{\mathbf{k} \in {\cal T}_3}
\widehat{D}_{N,\bf k}\,e^{i\mathbf{k \cdot x}},  $$
where, 
\BEQ
\label{eq:rep_D_N2}
 \begin{array}{l} \displaystyle
   \widehat{D}_{N,{\bf k}}=\sum_{n=0}^N\left(\frac{\alpha^{2p}|{\bf
         k}|^{2p}}{1+\alpha^{2p}|{\bf k}|^{2p}}\right)^n= (1+\alpha ^{2p} |{\bf
     k}|^{2p})\rho_{N,p,{\bf k}} , \\ 
     \hskip 7cm \displaystyle
     \rho_{N,p, {\bf
       k}}=1-\left(\frac{\alpha^{2p} |{\bf k}|^{2p}}{1+\alpha^{2p} |{\bf
         k}|^{2p}}\right)^{N+1}.
 \end{array}
\EEQ
The following holds \cite{BL11}:
 \begin{eqnarray}
 &&  \label{Prop-p1} 1\leq \widehat{D}_{N,\bf k} \leq N+1,   \quad  \forall\, {\bf k} \in {\cal T}_3, 
   \\
&&  \label{Prop-p2} \displaystyle \widehat{D}_{N,\bf k}) \approx (N+1)
   {1+\alpha^{2p} |{\bf k}|^{2p} \over \alpha^{2p} |{\bf k}|^{2p} }, \quad  \hbox{for
     large } |{\bf k}|,  
   \\
 && \label{Prop-p3} \displaystyle  \lim_{|{\bf k}
     |\to+\infty}\widehat{D}_{N,\bf k} =N+1,  \quad    \\   
&& \label{Prop-p4}  \widehat{D}_{N,\bf k}\leq(1+\alpha^{2p}|{\bf k}|^{2p}) = \widehat A_{\bf k},  \quad  \forall\,
   {\bf k} \in {\cal T}_3,  
 \end{eqnarray} 
where $\widehat A_{\bf k}$ is defined by (\ref{eq:transfer_inverse}). We deduce from (\ref{Prop-p1}) and (\ref{Prop-p3}):
\begin{lemma} \label{lem:D_N}
A real number $s \ge 0$ being given, the operator $D_N$ is a isomorphism over $  \mathbb{H}_s $, such that 
$1 \le || D_N || \le N+1$. Morover, the subspace of free divergence field ${\bf H}_s$ is stable under the action of $D_N$.   \fin
\end{lemma}


\subsection{Former Results} 

This section aims to recall results of \cite{BL11} about the system 
\begin{equation} \label{eq:adm1}
  \begin{aligned}
   \partial_t \moy_N +\nabla\cdot(\overline {D_{N} (\moy_N) \otimes
     D_{N} (\moy_N)})-\nu\Delta \moy_N +\nabla \pre_N = \overline \bff,   
   \\
   \nabla\cdot \moy_N =0,   
   \\
 \moy_N (0, \x) = \overline{\vit_0} (\x).
 \end{aligned}
\end{equation}
Throughout the paper, we assume that $\vit_0$ and ${\bf f}$ satisfy, 
\BEQ \label{eq:data} \vit_0 \in {\bf H}_0, \quad  {\bf f}\in  L^2([0,T] \times \tore)^3, \EEQ
and $\alpha>0$ is fixed. 

\begin{definition}[Regular Weak solution]\label{RegSol;p}
 We say that the couple $(\moy_N,\pre_N)$ is a ``regular weak solution" to
 system~(\ref{eq:adm1}) if and only if the three following items are
 satisfied: \medskip

\textcolor{red}{1) \underline{\sc Regularity} }

 \begin{eqnarray}
   && \label{Reg11-} \moy_N \in   L^2([0,T];
   \mathbf{H}_{1+p}) \cap C([0,T];  \mathbf{H}_{p}),
   \\
   && \label{Reg12-} \p_t \moy_N \in L^2([0,T];  \mathbf{H}_{0}) 
   \\
   && \label{Reg13-} \pre_N \in L^2([0,T]; H^1(\tore)),
 \end{eqnarray}

 \textcolor{red}{2) \underline {\sc Initial data} }
 \begin{equation}\label{Initial-}
   \displaystyle \lim_{t \rightarrow 0}\|\moy_N(t, \cdot) -  \overline{\vit_0}  \|_{ \mathbf{H}_{p}} = 0,
 \end{equation}

 \textcolor{red}{3) \underline{\sc Weak Formulation}}

 \begin{eqnarray}
   && \label{Reg14-} \forall \, \vittest \in L^2([0,T];
   H^1(\tore)^3), 
   \\ 
   &&  \label{RRQPJ-}
   \begin{aligned}
     \displaystyle\int_0^T\int_{\tore} \p_t \moy_N \cdot \vittest - 
     \int_0^T\int_{\tore} \overline{D_{N} (\moy_N) \otimes D_{N} (\moy_N)} : \g  \vittest + \nu 
     \int_0^T\int_{\tore} \g \moy_N : \g  \vittest \,  
     \\ 
     \displaystyle \hskip 6cm
     +\int_0^T\int_{\tore} \g \pre_N \cdot \vittest =\int_0^T\int_{\tore}  \overline {\bf f} \cdot \vittest.
   \end{aligned}
 \end{eqnarray}  \fin
\end{definition}
\begin{theorem}\label{thm:existence;p} (\cite{BL11}) Assume $p \ge 3/4$. 
  Problem (\ref{eq:adm1}) has a unique regular
 weak solution.  Moreover, when $p \ge 1$, 
 \BEQ 
 \label{eq:add_reg} 
\p_t \moy_N \in L^2([0,T], {\bf H}_{p-1}), \quad 
\pre_N  \in L^2([0,T],
{H}^{p}(\tore)). 
\EEQ \fin
\end{theorem}

\begin{theorem}~\label{thm:Principal-p} (\cite{BL11}) There exists a weak dissipative solution to the NSE (\ref{nse})
$$ (\vit, p) \in \left [ L^2 ([0,T], {\bf H}_1)  \cap L^2 ([0,T], {\bf H}_0) \right ] \times L^{5/3} ([0,T] \times \tore ) $$
such that from the sequence $(\moy_{N}, 
  \pre_{N})_{N \in \N}$, one can extract a sub-sequence (still denoted
  $(\moy_N, \pre_N)_{N \in \N}$) such that
\BEQ
   \begin{array}{ll}
  \moy_{N} \to \moy\quad& \left \{ \begin{array}{l} \text{weakly in }L^2 ([0,T], {\bf H}_{1+p}(\tore)^3) \cap
     L^\infty ([0,T], {\bf H}_p), \\ ~ \\
    \text{strongly in } L^r ([0,T];
     H^p(\tore)^3),\quad\forall \,1\leq r<+\infty, \end{array} \right. 
     \\
     ~
     \\
 \pre_{N}\to \pre &\text{weakly in }L^2([0,T];H^1(\tore)\cap
     L^{5/3}([0,T]; W^{2p,5/3}(\tore)),
   \end{array}
\EEQ \fin
 \ETHM

\section {Estimate of the modeling error}\label{sec:model_error}

\subsection{Regularity assumption and main result} \label{sec:reg_ass}

Let $(\moy_N, \pre_N)$ be the solution of Problem (\ref{eq:adm1}). We assume that  the limit $(\moy, \pre) = (G \vit, G p)$ of $(\moy_N, \pre_N)_{N \in \N}$ satisfies 
the regularity assumption
\BEQ \label{eq:reg_assump} \vit = A \moy \in L^4({\bf H}_1). \EEQ
By Sobolev injection Theorem, we deduce
\BEQ \label{eq:reg_assump_2}  \vit \in L^4 ([0,T] \times \tore). \EEQ
Since $(\vit, p)$ is solution to the NSE, one has 
\BEQ \Delta p = - \g \cdot ( \g \cdot ( \vit \otimes \vit) ) + \g \cdot {\bf f}, \EEQ 
which yields in the periodic case 
\BEQ \label{eq:reg_dt_p} p \in L^2 ([0,T]Ê\times \Omega), \EEQ
and we derive from the NSE, 
\BEQ \label{eq:reg_dt_u} \p_t \vit \in L^2([0,T], {\bf H}_{-1}). \EEQ
Our main result is 
\BTHM \label{thm:main} Let $\err_N = \moy - \moy_N$ be the error modeling, and assume that (\ref{eq:reg_assump}) holds. Then we have 
\BEQ \begin{array}{l}  \label{eq:cor_energy_final} \displaystyle ||\err_N (t, \cdot) ||_0^2 + \alpha^{2p} || \err_N (t, \cdot) ||_p^2 + 
\nu \int_0^t ( || \g \err_N (s, \cdot) ||_0^2 + \alpha^{2p} || \g \err_N (s, \cdot) ||_p^2) ds \le 
\\ \hskip 4cm \displaystyle \phantom{\sum_{k=0}^N}  { 16 C \alpha \,   \over   \nu ( 2p (N+1) )^{1/2p}  }  || \vit   ||^4_{L^4({\bf H}_1)} e^{{1 \over \nu^3} || \vit   ||^4_{L^4({\bf H}_1)}}. \end{array}
\EEQ
where $C$ is a universal constant, as a product of Sobolev constants. \footnote{For simplicity, we note $L^4({\bf H}_1)$ instead of $L^4([0,T], {\bf H}_1)$}  
\ETHM

\subsection{Modeling error and residual stress} 

Let $\err_N$ and $\res_N$ be the error modeling and the residual stress defined by 
\BEQ \begin{array} {rcl} \err_N &= &\moy - \moy_N, \\
\res_N & = & \vit \otimes \vit - D_N \moy\otimes D_N \moy. \end{array} \EEQ 
The equation satisfied by $\err_N$ is derived by substracting  (\ref{eq:adm1}) to the filtered NSE (\ref{smago}). Expressing the right hand side in terms of 
$\res_N$, we obtain 
\BEQ \label{eq:equation_for_err}  \p_t \err_N + \g \cdot (\overline{ D_N \err_N \otimes D_N \moy_N }) -\nu \Delta \err_N + \g r_N = - \g \cdot \overline \res_N - 
\g \cdot (\overline{D_N \moy \otimes D_N \err_N}),
\EEQ
where $r_N = \overline p - \pre_N$. 
\medskip

The aim of this section is to estimate $\err_N$ in terms of $\res_N$.  It adresses $A^{1/2}D_{N}^{1/2} \err_N $ rather than $\err_N$, since the natural 
multiplier to get an energy balance from equation (\ref{eq:equation_for_err}) is $AD_N \err_N$, and formally 
$(\p_t \err_N , AD_N \err_N) = (d / 2dt) || A^{1/2}D_N^{1/2} \err_N ||_0$. Once $A^{1/2}D_{N}^{1/2} \err_N$ is estimated, 
we derive bounds for 
$\err_N$ (Corollary \ref{cor:error_modeling} below) by comparing the norms of the various operators we consider. 

\marginpar{ \textcolor{red}{I found $4/ \nu$ instead of $8/\nu$ and $27/\nu$ in the exponential instead of $1/\nu$: this needs to be checked } }
\BTHM  \label{thm:estimate} The following inequality holds: 
\BEQ \label{eq:energy_err_mod}  \begin{array} {l} \displaystyle || A^{1/2}D_{N}^{1/2} \err_N (t, \cdot) ||_0^2  + \nu \int_0^t || A^{1/2}D_{N}^{1/2} \err_N (s, \cdot) ||_1^2 ds \le 
\\ \displaystyle
\hskip 7cm {8 \over \nu} e^{{1 \over \nu^3} || \vit   ||^4_{L^4({\bf H}_1)}} \int_0^t || \res_N (s, \cdot) ||^2_0 ds,  \end{array} \EEQ
 for all $N>0$ and $t\ge 0$. \fin
\ETHM

\proof The proof is based on an energy equality satisfied by $A^{1 \over 2}D_{N}^{1 \over 2} \err_N$ to which one applies 
Gronwall's Lemma. To do so, we use $AD_N \err_N$ as multiplier in  the (\ref{eq:equation_for_err}) satisfied by $\err_N$ and we integrate by parts.  
\medskip

The proof is divided into 
three steps. In a first one, we check that 
$AD_N \err_N$ is appropriate as multiplier to validate the procedure. In a second one,  we perform integrations by parts. In a last step, we apply 
usual interpolation inquality to be in order to apply Gronwall's Lemma. 

\medskip

\step {\sl Consistency of the procedure}. We check the regularity of $A^{1/2}D_{N}^{1/2} \err_N$  and each factor in equation (\ref{eq:equation_for_err}) one after each other, beginning with $\err_N$. 
The regularity assumption (\ref{eq:reg_assump}) combined with the regularization effect (\ref{eq:operator_G}) of operator G, gives 
$\moy \in L^4([0,T], {\bf H}_{1+2p})$. Therefore, we have at least by (\ref{Reg12-}) about $\moy_N$'s regularity,  
\BEQ \label{eq:reg_errN} \err_N \in L^2([0,T], {\bf H}_{1+p}) \subset  L^2([0,T],   \mathbb{H}_{1+p}),  \EEQ
where 
Applying Lemma \ref{lem:D_N} combined with  (\ref{eq:operator_G}), we get
\BEQ \label{eq:reg_ADNerrN} AD_N \err_N \in L^2([0,T] , \mathbb{H}_{1-p}). \EEQ 
We whish to prove now that each factor in equation  (\ref{eq:equation_for_err}) is at least in 
$$L^2([0,T] , \mathbb{H}_{p-1}) =  (L^2([0,T] , \mathbb{H}_{1-p}))' $$ 
(see subsection \ref{sec:space_functions}). To be synthetic, we write things as:
\BEQ \left. \begin{array} {rcr} (\ref{Reg12-}) &+ & (\ref {eq:reg_dt_u})
 \\ & +&  (\ref{eq:add_reg})  \end{array} \right \}  \Rightarrow   \left \{ \begin{array} {lr} \p_t \err_N \in L^2([0,T] , {\bf H}_{0}) & \hbox{if }Ê3/4 \le p \le 1, \\
   \p_t \err_N \in  L^2([0,T] ,  {\bf H}_{p-1}) & \hbox{if }Êp \ge 1. \end{array} \right. \EEQ
 When $3/4 \le p \le 1$, ${\bf H}_{0} \hookrightarrow  \mathbb{H}_0 \hookrightarrow  \mathbb{H}_{p-1}$, and when $p \ge 1$, 
  ${\bf H}_{p-1} \hookrightarrow  \mathbb{H}_{p-1}$. In all cases, 
  \BEQ \p_t \err_N \in  L^2([0,T] , \mathbb{H}_{p-1}). \EEQ
  Similarly,
\BEQ  \left. \begin{array} {rcr} (\ref{Reg13-}) &+ & (\ref {eq:reg_dt_p})
 \\ & +&  (\ref{eq:add_reg})  \end{array} \right \}  \Rightarrow   \left \{ \begin{array} {lr} \g r_N \in L^2([0,T] \times \tore )^3  & \hbox{if }Ê3/4 \le p \le 1, \\
   \g r_N \in  L^2([0,T] ,  H^{p-1}(\tore)^3) & \hbox{if }Êp \ge 1, \end{array} \right.  
\EEQ
that yields 
\BEQ \g r_N \in  L^2([0,T] , \mathbb{H}_{p-1}). \EEQ
From the injection ${\bf H}_1 \hookrightarrow \mathbb{H}_1$, we deduce
  \BEQ (\ref {eq:reg_errN})  \Rightarrow  \Delta \err_N \in L^2([0,T], \mathbb{H}_{p-1}). \EEQ
  Furthermore, as $(\vit, p)$ is a dissipative solution to the NSE, $\vit \in L^\infty([0,T], \mathbb{H}_0)$, therefore 
  $\moy \in L^\infty([0,T], \mathbb{H}_{2p})$, and by lemma \ref{lem:D_N}, we get
  \BEQ \label{eq:reg_DN_moy} D_N \moy \in L^\infty([0,T], \mathbb{H}_{2p}), \EEQ
from which we conclude
  \BEQ   \label{eq:reg_DN_errN} \left. \begin{array} {rcr} (\ref{Reg11-}) & +&  (\ref{eq:reg_DN_moy}) \\ 
 & + & \hbox{ lemma \ref{lem:D_N}} \end{array} \right \}  \Rightarrow D_N \err_N \in L^\infty([0,T], \mathbb{H}_{p}). \EEQ
Since $p \ge 3/4$, we deduce from Sobolev injection Theorem ${\bf H}_p \hookrightarrow L^4(\tore)^3$, that yields
\BEQ  \left. \begin{array} {rcr} (\ref{eq:reg_DN_moy}) & + & (\ref{eq:reg_DN_errN}) \\
& + & (\ref{eq:operator_G}) \end{array} \right \} \Rightarrow \g \cdot  ( \overline { D_N \err_N \otimes D_N \moy_N } ) \in L^\infty([0,T], \mathbb{H}_{2p-1}). \EEQ
Similarly, 
\BEQ  \g \cdot  ( \overline { D_N \moy \otimes D_N \err_N } ) \in L^\infty([0,T], \mathbb{H}_{2p-1}). \EEQ
Finally,  $\vit \in  L^\infty([0,T], \mathbb{H}_0)$ combined with
(\ref{eq:reg_assump_2}) and properties of $G$ and $D_N$ already mentioned, yields
\BEQ \g \cdot \overline \tau_N \in L^2([0,T], \mathbb{H}_{2p-1}). \EEQ
Bringing  together all these results, we conclude that when
$$ \mathbb{A}_N =   \\  \p_t \err_N + \g \cdot (\overline{ D_N \err_N \otimes D_N \moy_N }) -\nu \Delta \err_N + \g r_N + \g \cdot \overline \res_N+ 
\g \cdot (\overline{D_N \moy \otimes D_N \err_N})$$
then  $\mathbb{A}_N  \in L^2([0,T], \mathbb{H}_{p-1})$. Therefore, the duality pairing 
$_{p-1}( \mathbb{A}_N, AD_N \err_N)_{1-p}$, which makes consistent the multiplication of equation (\ref{eq:equation_for_err})  by 
$ AD_N \err_N$. In what follows, we omit the subscripts when writing duality pairings. 

 \medskip
 
\step {\sl Energy equality}.   
Since all the operators we consider are self adjoint, the following holds (see \cite{LM1968}):
\BEQ \begin{array} {l} \displaystyle  (\p_t \err_N, AD_N \err_N ) = {d \over 2 dt}|| A^{1 \over 2}D_N^{1 \over 2} \err_N ||_0^2, \\
(- \Delta \err_N, AD_N \err_N ) = || A^{1 \over 2}D_N^{1 \over 2} \err_N ||_1^2.
  \end{array} \EEQ
Furthermore, since $AD_N \err_N$ has zero divergence, $(\g r_N, AD_N \err_N ) = 0$. Finally, as the operators commute with the differential operators, 
\BEQ \label{eq:free_div} \begin{array} {l} (\g \cdot (\overline{ D_N \err_N \otimes D_N \wit_N }), AD_N \err_N) = (A^{-1} \g \cdot (D_N \err_N \otimes D_N \wit_N ), AD_N \err_N) = \\
(A^{-1} \g \cdot (D_N \err_N \otimes D_N \wit_N ), AD_N \err_N) = (\g \cdot (D_N \err_N \otimes D_N \wit_N ), D_N \err_N) = \\
\hfill ( ( D_N \wit_N \cdot \g ) D_N \err_N, D_N \err_N) = 0,  \end{array}  \EEQ
because $D_N \wit_N$ has zero divergence. Finally, arguing as in (\ref{eq:free_div}) to eliminate the bar in the integrals of right hand side, we get
\BEQ  \label{eq:energy_equality_1} {d \over 2 dt}|| A^{1 \over 2} D_N^{1 \over 2} \err_N ||_0^2 + \nu  || A^{1 \over 2}D_N^{1 \over 2} \err_N ||_1^2 = 
(\res_N, \g D_N \err_N) - ((D_N \err_N \cdot \g ) D_N \moy, D_N \err_N) \EEQ
\medskip

\step {\sl Bounds and Gronwal's lemma}. 
We bound each term of the right hand side of (\ref{eq:energy_equality_1}) after  each other. From Cauchy-Schwarz inequality combined with 
Young inequality, we get
\BEQ \label{eq:est_1}| (\res_N, \g D_N \err_N) | \le {4 \over \nu} ||\res ||_0^2 + {\nu \over 4} || D_N \err_N ||_1^2.  \EEQ
In the same way, by using Ladyzenskaya's inequality \cite{Tem2001} we obtain
\BEQ \label{eq:est_2} \begin{array} {l} | ((D_N \err_N \cdot \g ) D_N \moy, D_N \err_N) | \le || D_N \err_N ||_{L^4} ^2 || D_N \moy ||_1  \le \\
\hskip 7cm  || D_N \err_N ||_0^{1 \over 2} || D_N \err_N ||_1^{3 \over 2}|| D_N \moy ||_1 . \end{array} \EEQ
The symbol of $D_N G$ is equal to $\rho_{N,p,{\bf k}} \in [0,1]$ (see (\ref{eq:rep_D_N2})). Therefore, we have  $|| D_N \moy ||_1 \le || \vit ||_1$. By 
Young inequality combined with (\ref{eq:est_2}), we obtain
\BEQ  |((D_N \err_N \cdot \g ) D_N \moy, D_N \err_N) | \le {1 \over \nu^3} || \vit ||_1^4 || D_N \err_N ||_0^2 + {\nu \over 4} || D_N \err _N||_1^2.  \EEQ
We deduce  from (\ref{Prop-p4}) that the symbol of $D_N$ is less than the symbol of $A^{1/2}D_N^{1/2}$, which leads to
\BEQ \label{eq:est_3} || D_N \err_N ||_0 \le || A^{1 \over 2}D_{N}^{1 \over 2} \err_N||_0, \EEQ
regardless of $N$. Combining (\ref{eq:energy_equality_1}),  (\ref{eq:est_1}), (\ref{eq:est_2}) and  (\ref{eq:est_3}) yields 
\marginpar{ \textcolor{red}{I found here 
${4 \over \nu} || \res||_0^2 + {27 \over \nu^3} || \vit ||_1^4 || A^{1 \over 2}D_{N}^{1 \over 2} \err_N||_0^2 $ : TO BE CHECKED AGAIN} }
\BEQ \label{eq:eq_en_mo_er} {d \over  dt}|| A^{1 \over 2} D_N^{1 \over 2} \err_N ||_0^2 + \nu  || A^{1 \over 2}D_N^{1 \over 2} \err_N ||_1^2 \le 
{8 \over \nu} || \res||_0^2 + {1 \over \nu^3} || \vit ||_1^4 || A^{1 \over 2}D_{N}^{1 \over 2} \err_N||_0^2 \EEQ
Inequality (\ref{eq:energy_err_mod})  results from inequality (\ref{eq:eq_en_mo_er}) thanks to a standard generalisation of Gronwall's lemma \cite{TG19}.\hfill $\square$

\begin{Corollary} \label{cor:error_modeling} The error modeling $\err_N$ satisfies 
\BEQ \begin{array}{l}  \label{eq:cor_energy} \displaystyle ||\err_N (t, \cdot) ||_0^2 + \alpha^{2p} || \err_N (t, \cdot) ||_p^2 + 
\nu \int_0^t ( || \g \err_N (s, \cdot) ||_0^2 + \alpha^{2p} || \g \err_N (s, \cdot) ||_p^2) ds \le 
\\ \hskip 6,9cm \displaystyle {8  \over \nu} e^{{1 \over \nu^3} || \vit   ||^4_{L^4({\bf H}_1)}} \int_0^t || \res_N (s, \cdot) ||^2_0 ds, \end{array}
\EEQ
for all $N>0$ and $t\ge 0$. \fin
 \end{Corollary}

\proof Let $\displaystyle  \vittest  = \sum _{{\bf k}\in {\cal T}_3} \hat {\bf v}_{\bf k} e^{i {\bf k} \cdot \x} \in \mathbb{H}_p$. We observe that 
\BEQ  \label{eq:operators_inequality} ||A^{1 \over 2} {\bf v} ||_0 ^2 =  \sum _{{\bf k} \in {\cal T}_3} (1 + \alpha^{2p} |{\bf k} |^{2p})  |\hat {\bf v}_{\bf k}|^{2} = || {\bf v} ||_0^2 + \alpha^{2p}  || {\bf v} ||_p^2.\EEQ
We first  take ${\bf v} = D_N^{1/2} \err_N$ in (\ref{eq:operators_inequality}). By using   (\ref{Prop-p1}), which yields the general formal inequality 
$|| \wit ||_s \le || D_N^{1 /2} \wit ||_s$, we deduce the further inequality
\BEQ \label{eq:operators_inequality_1} || \err_N ||_0^2 + \alpha^{2p} || \err_N ||_{p}^2 \le || A^{1/2} D_N^{1/2}Ê\err_N ||_0^2. \EEQ
We next take  ${\bf v} = \p_i D_N^{1/2} \err_N$ in (\ref{eq:operators_inequality}), which yields
\BEQ \label{eq:operators_inequality_2} || \g  \err_N ||_0^2 + \alpha^{2p} || \g  \err_N ||_{p}^2 \le || A^{1/2} D_N^{1/2}Ê\err_N ||_1^2. \EEQ 
We deduce (\ref{eq:cor_energy}) from (\ref{eq:energy_err_mod}) thanks to (\ref{eq:operators_inequality_1}) and  (\ref{eq:operators_inequality_2}). 

 \fin

\section{Residual stress and rate of convergence} \label{sec:res_stres}\label{sec:res_stress}

Now that we have shown that the modeling error  $\err_N$ is driven by the $L^2$ norm of the residual stress $\res_N$, involving the $L^4({\bf H}_1)$ norm of $\vit$, it remains estimate the $L^2$ norm of $\res_N$, which what we aim to carry out in this section. Framework, assumptions and notations are those of section \ref{sec:model_error}. 
\medskip

In what follows, $S_s$ denotes the Sobolev constant\footnote{The constants $S_1$ and $S_{1/2}$ do not depend on $L$. One can prove that $S_1 \le (16 + 3/\pi)^{1/3}$, see \cite{RLbookNSE}. Unfortunately, we do not know any numerical bound for $S_{1/2}$, even such a bound may probably be found in the litterature} in the injection $ \mathbb{H}_s \hookrightarrow L^{s^\star} (\tore)^3$.
To begin with, we show 

\begin{lemma}Ê\label{lem:residual_stress}  The following inequalities holds true
\begin{eqnarray} && \label{eq:res_stress_est_princ} || \res_N  ||^2_0 \le 2 C \,  || \vit (t, \cdot) ||_1^2 \,  || \vit - D_N \moy   ||^2_{1/2}, \\
&&  \label{eq:fin} || \vit - D_N\moy  ||^2_{1/2} \le  { \alpha\,   \over  ( 2p (N+1))^{1 / 2p}Ê } ||\vit ||_1^2,  \phantom{\int_0^1}\end{eqnarray} 
where $C = S_1 S_{1/2}$.\footnote{Inequalities (\ref{eq:res_stress_est_princ}) and (\ref{eq:fin}) both hold at any fixed time 
$t \in [0,T]$, which is not indicated here to reduce the notations.}  \fin
\end{lemma} 

\proof \step \label{step:etap1_res_stress} {\sl Proof of (\ref{eq:res_stress_est_princ}).  } We write $\res_N$ as 
\BEQ \res_N = (\vit- D_N \moy ) \otimes \vit + D_N \moy \otimes (\vit- D_N \moy ). \EEQ
Therefore, combining H\"older inequality with $1/3+1/6 = 1/2$ for conjugation, to the Sobolev inequality $ ||\wit ||_{L^6} \le  S_1 || \wit ||_1$, we get
\BEQ \label{eq:res_stress_est1} || \res ||^2_0 \le 2 S_1 || \vit ||_1^2 || \vit - D_N \moy ||^2_{L^3(\tore)^3},\EEQ
To estimate $|| \vit - D_N \moy ||^2_{L^3(\tore)^3}$,  we use the injection of $ \mathbb{H}_{1/2}$ onto $L^3(\tore)^3$ to obtain
\BEQ \label{eq:estim} || \vit - D_N \moy ||^2_{L^3(\tore)^3} \le S_{1/2}  || \vit - D_N \moy ||^2_{1/2}, \EEQ
hence (\ref{eq:res_stress_est_princ}) by combining  (\ref{eq:res_stress_est1})  and (\ref{eq:estim}). \fin

\medskip

\step \label{step:etap2_res_stress} {\sl Proof of (\ref{eq:fin}).  } We deduce from (\ref{eq:rep_D_N2}), 
\BEQ \label{dec_norm}|| \vit - D_N \moy ||^2_{1/2} =  \sum _{ {\bf k}\in {\cal T}_3 } \left ( { \alpha^{2p} |{\bf k} |^{2p}  \over 1+ \alpha^{2p} |{\bf k} |^{2p} } \right )^{2(N+1)} |{\bf k}| \,  | \hat {\bf u}_{\bf k}|^2, \EEQ 
 We apply the technical inequality (\ref{eq:inq_tech1}) proved in Appendix 
\ref{sec:appendix} below, with $x=\alpha^p |{\bf k}
  |^p$, $a=2p(N+1) >1$, $b=0$, which yields 
\BEQ \label{eq:ine_dec1} \left ( { \alpha^{2p} |{\bf k} |^{2p}  \over 1+ \alpha^{2p} |{\bf k} |^{2p} } \right )^{2p(N+1)} \le {\alpha^p | {\bf k} |^p \over \sqrt{2p (N+1) }}  . \EEQ
We raise both sides of (\ref{eq:ine_dec1}) to the power $1/p$, we
multiply the result by $| {\bf k}| | \hat
{\bf u}_{\bf k}|^2$ and get
\BEQ  \left ( { \alpha^{2p} |{\bf k} |^{2p}  \over
  1+ \alpha^{2p} |{\bf k} |^{2p} } \right )^{2(N+1)} |{\bf k} | | \hat
{\bf u}_{\bf k}|^2\le
     {\alpha \over (2p (N+1) )^{1/2p} }  | {\bf k} | ^2| \hat
{\bf u}_{\bf k}|^2, \EEQ
hence  (\ref{eq:fin}) from (\ref{dec_norm}).

\begin{corollary}  \label{thm:principal_2} The following estimate holds
\BEQ \label{eq:est_res2}  || \res_N (t, \cdot) ||_0^2 \le  { 2 C \alpha \,   \over   ( 2p (N+1) )^{1/2p}  } ||\vit (t, \cdot) ||_1^4, \EEQ
for all $t \in [0,T]$. 
\end{corollary} 
Inequality (\ref{eq:est_res2}) results from (\ref{eq:fin}) combined to (\ref{eq:res_stress_est_princ}).  \hfill $\square$

\medskip
Summarizing: $(\ref{eq:cor_energy})+ (\ref{eq:est_res2}) \Rightarrow$ 
\BEQ \begin{array}{l}  \label{eq:cor_energy_final2} \displaystyle ||\err_N (t, \cdot) ||_0^2 + \alpha^{2p} || \err_N (t, \cdot) ||_p^2 + 
\nu \int_0^t ( || \g \err_N (s, \cdot) ||_0^2 + \alpha^{2p} || \g \err_N (s, \cdot) ||_p^2) ds \le 
\\ \hskip 4cm \displaystyle \phantom{\sum_{k=0}^N}  { 16 C \alpha \,   \over   \nu ( 2p (N+1) )^{1/2p}  }  || \vit   ||^4_{L^4({\bf H}_1)} e^{{1 \over \nu^3} || \vit   ||^4_{L^4({\bf H}_1)}}. \end{array}
\EEQ
for all $N>0$ and $t\ge 0$.

\section{Case of Gaussian filter}Ê

\subsection{Framework} \label{sec:framework_gauss} 

The Gaussian filter is specified by its kernel, 
\BEQ \label{eq:gauss_filter} 
\tilde G_{\alpha}(\bx)   = \tilde G (\x) =  \left(\frac{6}{\alpha^2\pi}\right)^{3/2} \exp
\left(-\frac{6}{\alpha^2}|| \x||^2\right), 
\EEQ  
where we omit the subscript $\alpha$ for simplicity. It can be shown that \cite{SS03}, 
\BEQ \tilde G(\bx)  = \sum_{\mathbf{k} \in {\cal T}_3}  \tilde G_{\bf k}e^{i\mathbf{k \cdot x}} \phantom{\int_0^1}
\quad \hbox{where}Ê\quad 
\tilde {G}_{\bf k} = e^{-\frac{\alpha^{2}|\mathbf{k}|^{2}}{24}}
\EEQ
Let $s \ge 0$ and $q \ge s$. There exists a constant $C$ be such that 
\BEQ \forall \, {\bf k}Ê\in {\cal T}_3^\star, \quad \tilde G_{\bf k} |  {\bf k} |^q \le C | {\bf k} |^s. \EEQ
Therefore,  
 \BEQ \forall \, s \ge 0, \quad \forall \,   {\bf u}Ê\in \esp_s, \quad \forall q \ge s, \quad  \tilde G {\bf u} \in \esp_q. \EEQ 
Let  ${\bf u} $ being given such that $\forall \, {\bf k}Ê\in {\cal T}_3^\star$, $|Ê\hat {\bf u} _{\bf k}Ê|Ê= |{\bf k} |^{-1-q} \not=0$ ($q \ge 0$). Such 
a vector field ${\bf u}$ belongs to $\esp_q$, but it easy checked that $\tilde G^{-1} \notin \esp_s$ for any $s$. This is why the theory above about Helmholz filters  fails, since it is based 
on the fact that $G$ defines an isomorphism between $\esp_s$ spaces. 
\medskip

However, ADM may be considered for the Gaussian filter, and the resulting model yields a well posed problem \cite{stanculescu}.  Moreover, we shall show in what follows that 
it can be approached in some sense, by a sequence of operators which fall within the framework of the theory exposed above. 

\subsection {Approximation of the Gaussian filter}Ê

we note that 
for all ${\bf k}Ê\in {\cal T}_3^\star$ fixed, 
\BEQ \tilde {G}_{\bf k} = \lim_{m \rightarrow}  \tilde{G}_{m,\bf k}, \quad \hbox{where} \quad 
 \tilde{G}_{m,\bf k} =\left (1+\frac{\alpha^{2}|\mathbf{k}|^{2}}{24m} \right )^{-m}
\EEQ
Let $\tilde G_m$ denotes the kernel 
\begin{equation} \label{eq:gauss_approx2} \displaystyle \tilde G_{m}(\x)  =\sum_{\mathbf{k} \in {\cal T}_3}
\left (1 + {\alpha ^{2} |{\bf k}|^{2} \over 24 m} \right )^{-m} e^{i\mathbf{k \cdot x}},  \end{equation}
which corresponds to the operator, still denoted by $\tilde G_{m}$,  
\BEQ \label{eq:Gm}  \tilde G_{m} = \left (1 - {\alpha ^{2}  \over 24 m } \Delta \right )^{-m} . \EEQ
In a sense that needs to be precised, the sequence $(\tilde G_{m} )_{m \in \N}$ converges to $\tilde G$. To be more specific, 
\begin{lemma}ÊFor all ${\bf k} \in {\cal T}_3^\star$, 
\begin{equation}\label{transf_est1}
| \tilde {G}_{\bf k} -  \tilde{G}_{m,\bf k} | \le {2 \over m}.\footnote{This estimate is uniform in ${\bf k}$, but unfortunately we cannot  conclude from this the normal convergence 
of the kernel sequence $(\tilde G_{m} )_{m \in \N}$ because the serie $1/m$ is not convergent}
\end{equation}
\end{lemma} 

\proof We prove in 
Appendix \ref{sec:appendix} the technical inequality (\ref{transf_est}), 
\[  \forall \, x \ge 0, \quad \forall m \ge 1, \quad 
\left| \left (1+{x \over m} \right )^{-m}-e^{-x} \right| \leq \frac{2}{m}.
\]
We deduce inequality (\ref{transf_est1}) in  replacing in this inequality $x$ by 
$\displaystyle \frac{\alpha^{2}|\mathbf{k}|^{2}}{24}$. \fin
\medskip

The following corollary is straightforward: 
\begin{corollary} \label{cor:cor_gauss}  For all $\bf u \in \mathbb{H}_s$, 
\BEQ || \tilde G {\bf u}  - \tilde G_m {\bf u} ||_s \le {2 \over m}Ê|| {\bf u}Ê||_s. \EEQ
\end{corollary} 
In other words, there is weak star convergence of the sequence of operators $(\tilde G_{m} )_{m \in \N}$ to the Gaussian filter 
$\tilde G$ in $\esp_s$ ($s \ge 0$). 

\subsection{Powers of the second order filter} \label{sec:powers_second_order} 

In what follows, we put for $m$ fixed,
\BEQ  \label{eq:mu_alpha} \mu^2 = {{\alpha^2}Ê\over 24 m}. \EEQ
and we denote by $H_m$  the 
$m^{\hbox{\tiny th}}$ power of the second order Helmholz operator
\BEQ H_m = ({\rm I} - \mu ^2 \Delta)^{-m}.  \EEQ
 Estimating the error modeling that corresponds to $H_m$ yields estimates for the error modeling that corresponds to 
$G_m$. The theroy developed above about Helmholz operators applies to operator $H_m$. Indeed, let 
\BEQ \hat H_{m, {\bf k} } = { 1 \over  (1 + \mu^2 | {\bf k} |^2)^m } \EEQ
be the symbol of $H_m$.  
Using the scalar inequality $1+x^m\leq (1+x)^m \leq 2^{m-1}(1+x^m)$
for positive $x$, we get
\begin{equation} \label{eq:helm_gen} 
 \frac{1}{2^{m-1}(1+\mu^{2m}|\mathbf{k}|^{2m})}\leq \hat{H}_{m,{\bf k}} \leq  \frac{1}{1+\mu^{2m}|\mathbf{k}|^{2m}}. 
\end{equation}
Using results of \cite{BL11} (section 6), we deduce from (\ref{eq:helm_gen}) that the ADM corresponding to 
$H_m$ has a unique regular weak solution $(\moy_{N}, \pre_{N})$, in the meaning of Definition \ref{RegSol;p} with $p=m$. Furthermore, this sequence of solution converges to some 
$(\moy, \pre)$ solution of the filtered NSE when $N$ goes to infinity. Thus we can perform the programme to estimate $\err_N$ in this case. We next prove. 

\BTHM \label{thm:main2} Let $\err_N = \moy - \moy_N$ be the error modeling corresponding to $H_m$, and assume that (\ref{eq:reg_assump}) still holds. Then we have \footnote{The constant $C$ below is as in Theorem \ref{thm:main}, inequality (\ref{eq:energy_err_mod}), see also Lemma \ref{lem:residual_stress}.  } 
\BEQ \begin{array}{l}  \label{eq:cor_energy_final2} \displaystyle ||\err_N (t, \cdot) ||_0^2 + \mu^{2m} || \err_N (t, \cdot) ||_m^2 + 
\nu \int_0^t ( || \g \err_N (s, \cdot) ||_0^2 + \mu^{2m} || \g \err_N (s, \cdot) ||_m^2) ds \le 
\\ \hskip 4cm \displaystyle \phantom{\sum_{k=0}^N}  { 14 C \mu \, m^{1 / 2}  \over   \nu ( 4 (N+1) )^{1/2m}  }  || \vit   ||^4_{L^4({\bf H}_1)} e^{{1 \over \nu^3} || \vit   ||^4_{L^4({\bf H}_1)}}.  \end{array} 
\EEQ \fin
\ETHM
\proof Thanks to (\ref{eq:helm_gen}), one can copy line by line proofs  of Theorem (\ref{thm:estimate}) and Corollary (\ref{cor:error_modeling}) and 
derive
\BEQ \begin{array}{l}  \label{eq:cor_energy2} \displaystyle ||\err_N (t, \cdot) ||_0^2 + \mu^{2m} || \err_N (t, \cdot) ||_m^2 + 
\nu \int_0^t ( || \g \err_N (s, \cdot) ||_0^2 + \mu^{2m} || \g \err_N (s, \cdot) ||_m^2) ds \le 
\\ \hskip 6,9cm \displaystyle {8  \over \nu} e^{{1 \over \nu^3} || \vit   ||^4_{L^4({\bf H}_1)}} \int_0^t || \res_N (s, \cdot) ||^2_0 ds. \end{array}
\EEQ
It remains to estimate $|| \res_N (s, \cdot) ||^2_0$. Step \ref{step:etap1_res_stress}  Êin the proof of Lemma \ref{lem:residual_stress} can be recycled, so that (\ref{eq:estim}) still holds in this case.   Therefore, we only have to bound 
\BEQ || \vit - D_N \moy ||^2_{1/2} =  \sum _{ {\bf k}\in {\cal T}_3^\star } \left(1- \frac{1}{\left(1+\mu^{2} |{\bf k} |^{2}\right)^m}\right)^{2(N+1)} |{\bf k}| \,  | \hat {\bf u}_{\bf k}|^2, \EEQ 
where as usual $\vit = \sum _{{\bf k}\in {\cal T}_3^\star} \hat {\bf u}_{\bf k} e^{i {\bf k} \cdot \x}$. We apply the technical inequality (\ref{eq:inq_tech3}) proved in Appendix 
\ref{sec:appendix} below, with $x=\mu |{\bf k}
  |$, $a=2(N+1) >1$, $m\geq 1$.
 We obtain 
\BEQ \label{eq:ine_dec} \left(1- \frac{1}{\left(1+\mu^{2} |{\bf k}|^{2}\right)^m}\right)^{2(N+1)} \le  {\sqrt{m} \, \mu \over (4 (N+1) )^{1/2m}} | {\bf k} |   . \EEQ
We multiply the result by $| {\bf k}| | \hat {\bf u}_{\bf k}|^2$ and get
\BEQ \label{eq:ine_dec} \left(1- \frac{1}{\left(1+\mu^{2} |k|^{2}\right)^m}\right)^{2N+2} |{\bf k} | | \hat
{\bf u}_{\bf k}|^2 \le  {\sqrt{m}\, \mu \over (4 (N+1))^{1/2m}}  | {\bf k} | ^2| \hat
{\bf u}_{\bf k}|^2,\EEQ
hence 
\BEQ  || \vit - D_N \moy ||^2_{1/2} \le {\sqrt{m}\, \mu \over (4 (N+1))^{1/2m}} || \vit ||_1, \EEQ
which  yields by (\ref{eq:estim}),
\BEQ \label{eq:res_stress_est} || \res ||^2_0 \le {2\sqrt{m} \, \mu \over (4 (N+1) )^{1/2m}} || \vit ||_1^4,\EEQ
giving (\ref{eq:cor_energy_final2}) thanks to (\ref{eq:cor_energy2}). \fin

\subsection{Passing to the limit} 

From the results of subsection \ref{sec:powers_second_order}, we deduce thanks to the relation (\ref{eq:mu_alpha}) 
 that the ADM  associated  to  the
filter specified by (\ref{sec:powers_second_order}), has a unique solution 
$(\moy_{N, m}, \pre_{N,m})$ which converges to some solution $(\moy_m, \pre_m)$, of the filtered NSE, by assuming that $(\vit_m, p_m)$ satisfies the regularity assumption (\ref{eq:reg_assump}).  
\medskip

Let $\err_{N, m} = \moy_m - \moy_{N, m} $ denotes the corresponding error modeling. Thanks to   (\ref{eq:cor_energy_final2}), we obtain\footnote{For the simplicity, we use $\mu m^{1/2} \le 5 \alpha$ instead of   (\ref{eq:mu_alpha}) } 
\BEQ  \label{eq:final_bound} \begin{array}{lc}   \displaystyle ||\err_{N, m} (t, \cdot) ||_0^2 + \alpha^{2m} (24 m)^{-m}  || \err_{N, m} (t, \cdot) ||_m^2 &  +  \\
 \displaystyle \phantom{\sum_{k=0}^N}    \nu \int_0^t ( || \g \err_{N, m} (s, \cdot) ||_0^2  &  +   \\  
\phantom{\sum_{k=0}^N}   \alpha^{2m} (24 m)^{-m} || \g \err_{N, m} (s, \cdot) ||_m^2) ds & \le  \\
 \hskip 4cm \displaystyle  \phantom{\sum_{k=0}^N}  { 70 C \alpha  \over   \nu ( 4 (N+1) )^{1/2m}  }  || \vit   ||^4_{L^4({\bf H}_1)} e^{{1 \over \nu^3} || \vit   ||^4_{L^4({\bf H}_1)}}.  \end{array}
\EEQ
Without any convergence result about ADM's associated to Gaussian filter  (\ref{eq:gauss_filter}) when $N$ goes to infinity, we cannot   consider the corresponding 
error modeling, and therefore take the limit in  (\ref{eq:final_bound}) when $m$ goes to infinity. Nervertheless, we observe that for a fixed $N$, the r.h.s of 
(\ref{eq:gauss_filter}) converges, as $m \rightarrow \infty$, to some $C=C(\nu, \alpha, \vit, C)$, which do not depend on $N$. We only 
can deduce a bound about the sup limit of the terms in the r.h.s.

\section{Conclusions and open problems} 

\subsection{Typical size of the constants}  

The main estimate (\ref{eq:cor_energy_final}) we get in the paper yields the rate of convergence to zero of the order modeling in the case of Helmholz filter of order 
$p$. The bound involes a constant  of the form
\BEQ  \kappa = {1 \over \nu}  || \vit   ||^4_{L^4({\bf H}_1)} e^{{1 \over \nu^3} || \vit   ||^4_{L^4({\bf H}_1)}}. \EEQ
The number of iteration $N$ requiered to reduce substancially the SFS area is driven by the size of the constant 
$\kappa$. 

\medskip 

This constant involves gradients of the true velocity of the fluid, which may be huge. For instance, in some turbulent boundary layer, one may observe 
flows for which $\g \vit$ is of order $3. 10^{4}$ $s^{-1}$ in layers thick of about $10^{-1}$ $m$. For such a air layer at $50^\circ$  (that can be considered as incomrpessible) of width and length equal to $1$ $m$, over a time range 
of $1$ $s$, with $\nu \approx 20. 10^{-6} $ $m^2 s^{-1}$, we find 
$$ \kappa \approx 10^{ 10^{28} } \, m^4 s^{-2},$$ 
which is a very huge constant. Therefore, even if the resolution would be of order $\alpha = 10^{-18}$ $m$, to fully  solve such a flow, the number of 
iteration $N$ required  to substancially reduce $\err_N$ is so large that the deconvolution algorithm seems not suitable for practical simulations, which is in contradiction with  results of \cite{SAK2001}, suggesting  that very few iterations are sufficient to significantly  reduce  the SFS area. 

\medskip

The rate of convergence as $ (p (N+1))^{-1/4p}$ comes from estimating norms of the residual stress $\tau_N$ involved in the equation 
for $\err_N$, whereas the constant $\kappa$ considered above comes from Gronwall's Lemma, which is known to lead to non optimal results. This yields the conjecture 
that the rate of convergence we found is optimal, which is not the case of the constant, that might be substancially improved. Furthermore, 
how the regularity assumption $\vit \in L^4({\bf H}_1)$ could be prevented ? 

\subsection{Gaussian Filter} 

It also remains the issue of convergence of ADM in the case of Gaussian filter. We conjecture that the convergence holds, but in a very weak sense, 
according to Corollary \ref{cor:cor_gauss}, a weak sense as yet undefined.

\section{Appendix} \label{sec:appendix}

This technical appendix aims at proving a general inequality that has been used in the proof of the estimate (\ref{eq:est_res2}). The result is the following.

\begin{theorem}\label{real_in}
The scalar inequality

\BEQ \label{eq:inq_tech2} 
\left(1-\frac{1}{(1+x)^m }\right)^a \leq \frac{mx}{\sqrt[m]{a}}
\EEQ

holds true for any $x\geq 0$, $a,m\geq 1$. \fin
\end{theorem}

\noindent We consider the LHS  function $$h(x)=\left(1-\frac{1}{(1+x)^m
}\right)^a $$ and fixed parameters $a,m\geq 1$.

\medskip

Its derivative is

$$
h'(x)=am \left(1-\frac{1}{(1+x)^m }\right)^{a-1}\frac{1}{(x+1)^{m+1}}
$$

We apply to $h(x)$ the Lagrange intermediate formula on $[0,x]$ and
get

$$
\left(1-\frac{1}{(1+x)^m }\right)^a =(x-0)h'(\psi)=xam \left(1-\frac{1}{(1+\psi)^m }\right)^{a-1}\frac{1}{(\psi+1)^{m+1}}
$$

for some $\psi\in (0,x)$.

\medskip

The inequality becomes

$$
xam \left(1-\frac{1}{(1+\psi)^m }\right)^{a-1}\frac{1}{(\psi+1)^{m+1}} 
\leq \frac{mx}{\sqrt[m]{a}}
$$

i.e.(after reducing $xm$ from both sides)

$$
a \left(1-\frac{1}{(1+\psi)^m }\right)^{a-1}\frac{1}{(\psi+1)^{m+1}} 
\leq \frac{1}{\sqrt[m]{a}}
$$

So now it's enough to prove that

$$
a \left(1-\frac{1}{(1+x)^m }\right)^{a-1}\frac{1}{(x+1)^{m+1}} 
\leq \frac{1}{\sqrt[m]{a}}
$$

for any $x\geq 0$, $a,m\geq 1$.

\medskip

To easy computations we make the substitution 

$$
y=\frac{1}{1+x} \in (0,1)
$$

and the inequality becomes

$$
a \left(1-y^m \right)^{a-1}y^{m+1} 
\leq \frac{1}{\sqrt[m]{a}}
$$

or, after putting $a$ on the RHS

$$
\left(1-y^m \right)^{a-1}y^{m+1}
\leq a^{-1-1/m }
$$

for any $y\in (0,1)$, $a,m\geq 1$.

\medskip

We denote the LSH above by  $g(y)=\left(1-y^m \right)^{a-1}y^{m+1} $

\medskip

Its derivative with respect to $y$ is

$$
g'(y)=-y^m (1-y^m)^{a-2}\left((a m +1)y^m   - (m + 1)\right)
$$

We see that the derivative vanishes at  
\[
y_0=\left(\frac{m + 1}{a m +1}\right)^{1/m}
\]

and is first positive on $[0, \left(\frac{m + 1}{a m +1}\right)^{1/m}]$ then negative on $[\left(\frac{m + 1}{a m +1}\right)^{1/m}, 1]$
therefore the maximum of $g$ is attained at $\left(\frac{m + 1}{a m +1}\right)^{1/m}$ and is equal to

\[
g(\left(\frac{m + 1}{a m +1}\right)^{1/m})= \left(1- \frac{m + 1}{a m +1}\right)^{a-1}\left(\frac{m + 1}{a m +1}\right)^{\frac{m+1}{m}}=
 \left(\frac{am -m}{a m +1} \right)^{a-1}\left(\frac{m + 1}{a m +1}\right)^{\frac{m+1}{m}}
\]

So now we need to show that

\[
 \left(\frac{am -m}{a m +1} \right)^{a-1}\left(\frac{m + 1}{a m +1}\right)^{\frac{m+1}{m}}\leq a^{-1-1/m }
\]
for $a,m\geq 1$.

\medskip

Now polish a bit the formula above. In the first term on LHS we simplify m, 

\[
 \left(\frac{a -1}{a  +1/m} \right)^{a-1}\left(\frac{m + 1}{a m +1}\right)^{\frac{m+1}{m}}\leq a^{-1-1/m }
\]

 In the bottom of the second term on LHS we pull out  a, 

\[
 \left(\frac{a -1}{a  +1/m} \right)^{a-1}  a^{-1-1/m} \left(\frac{m + 1}{ m +1/a}\right)^{\frac{m+1}{m}}\leq a^{-1-1/m }
\]

then cancel $a^{-1-1/m }$

\[
 \left(\frac{a -1}{a  +1/m} \right)^{a-1}  \left(\frac{m + 1}{ m +1/a}\right)^{\frac{m+1}{m}}\leq 1
\]

Then  in the second term on LHS we simplify m, 

\[
 \left(\frac{a -1}{a  +1/m} \right)^{a-1}  \left(\frac{1+ 1/m}{ 1 +1/(ma)}\right)^{\frac{m+1}{m}}\leq 1
\]

Now let $z=1/m\in (0,1]$

\medskip

The inequality becomes

\[
 \left(\frac{a -1}{a  +z} \right)^{a-1}  \left(\frac{1+ z}{ 1 +z/a}\right)^{1+z}\leq 1
\]

for any $a\geq 1$, $z\in  (0,1]$.

\medskip

We apply the natural log to both sides. We need to show that

\[
 (a-1)ln(a-1)-(a-1)ln(a+z)  + (1+z)ln(1+z)-(1+z)ln(1+z/a) \leq 0
\]

for any $a\geq 1$, $z\in  (0,1]$.

\medskip

Let $$f(a)= (a-1)ln(a-1)-(a-1)ln(a+z)  + (1+z)ln(1+z)-(1+z)ln(1+z/a) $$ be the LHS in the inequality above as a function of a.

\medskip

The derivative of f (with respect to a)

\[
f'(a)=ln(a-1)-ln(z+a)+\frac{1+z}{a}
\]

The second derivative is

\[
f''(a)=-\frac{(z+1)(az-z-a)}{a^2(a-1)(z+a)}
\]

Obviously, since $a\geq 1$, $z\in [0,1)$ we have that $az-a \leq 0$, so $az-z-a \leq 0$

therefore

\[
f''(a)=-\frac{(z+1)(az-z-a)}{a^2(a-1)(z+a)} \geq 0
\]

We conclude that the first derivative is increasing, therefore 

\[
f'(a)\leq \underset{a\to \infty}{lim} f'(a)=\underset{a\to \infty}{lim} \left(ln(a-1)-ln(z+a)+\frac{1+z}{a}\right) =\underset{a\to \infty}{lim}\left( ln\left(\frac{a-1}{z+a}\right)+\frac{1+z}{a}\right)=0
\]

Therefore $f'$ is negative, so f is decreasing. It follows that

\[
f(a)\leq\underset{a\to 1}{lim} f(a)=\underset{a\to 1}{lim} \left( (a-1)ln(a-1)-(a-1)ln(a+z)  + (1+z)ln(1+z)-(1+z)ln(1+\frac{z}{a}) \right)
\]

We know that in general \[ \underset{x\to 0}{lim} \,\,xln(x) =0\] therefore, the above limit is zero

\[
\underset{a\to 1}{lim}  (a-1)ln(a-1)-(a-1)ln(a+z)  + (1+z)ln(1+z)-(1+z)ln(1+\frac{z}{a})  = \]
\[=0-0+(1+z)ln(1+z)-(1+z)ln(1+z)=0
\]

We conclude that $f(a)\leq 0$ which proves the inequality. \fin

\begin{corollary}\label{real_in}
The scalar inequality

\BEQ  \label{eq:inq_tech3} 
\left(1-\frac{1}{(1+x^2)^m }\right)^a \leq \frac{\sqrt{m}x}{\sqrt[2m]{2a}}
\EEQ

holds true for any $x\geq 0$, $a,m\geq 1$.
\end{corollary}

In the previous inequality we replace $x$ with $x^2$ and get

\BEQ 
\left(1-\frac{1}{(1+x^2)^m }\right)^a \leq \frac{mx^2}{\sqrt[m]{a}}
\EEQ

for any $x\geq 0$, $a,m\geq 1$.

Replace in this inequality $a$ with $2a$, still keep $a\geq 1$ (but works for $a\geq 1/2$)

\BEQ 
\left(1-\frac{1}{(1+x^2)^m }\right)^{2a} \leq \frac{mx^2}{\sqrt[m]{2a}}
\EEQ

for any $x\geq 0$, $a,m\geq 1$.

Now extract the square root of both sides

\BEQ 
\left(1-\frac{1}{(1+x^2)^m }\right)^{a} \leq \frac{\sqrt{m}x}{\sqrt[2m]{2a}}
\EEQ

\begin{remark}
Setting $m=1$ in the previous 
inequality gives

\BEQ \label{eq:inq_tech1} 
\left(\frac{x^2}{1+x^2 }\right)^a \leq \frac{x}{\sqrt{2a}}\leq \frac{x}{\sqrt{a}}
\EEQ
for any $x\geq 0$, $a\geq 1$.

\end{remark}

\medskip

The following inequality will be used to approximate the Gaussian
filter with a power of the second order Helmholz filter and calculate
the accuracy of this approximation.

\begin{theorem}\label{transfer_est}
The scalar inequality

\BEQ \label{transf_est} 
\left| (1+x/n)^{-n}-e^{-x} \right| \leq \frac{2}{n}
\EEQ

is valid for any real $x\geq 0$ and any integer $n\geq 1$.
\end{theorem}

It is well-known that as a function of $n$ (and fixed $x\geq 0$) the
expression

\[
(1+x/n)^{-n}
\] 
is decreasing and converges to $e^{-x}$ as $n\to \infty$.
\textcolor{red}{ (this is elementary calculus, i ommited the proof.)
}

\medskip

Therefore, the left hand side in (\ref{transf_est})
can be written as

\[
\left | (1+x/n)^{-n}-e^{-x} \right |= (1+x/n)^{-n}-e^{-x}=e^{-n ln(1+x/n)}-e^{-x}
=e^{-n ln(1+y)}-e^{-n y}\]

where $y=x/n\geq 0$ .

\medskip

Applying the intermediate value theorem of Lagrange (corresponding to
the function $\xi\to e^{-n\xi}$) to the last term
above we get that

\[
e^{-n ln(1+y)}-e^{-n y} = n e^{-n \xi } (y-ln(1+y)) 
\]

for some $\xi\in [ln(1+y),y] $. Here we used $ln(1+y)\leq y$ for
$y\geq 0$. 

Since $e^{-n \xi } \leq e^{-n ln(1+y)}$
 we further have that

\begin{equation} \label{inq1}
e^{-n ln(1+y)}-e^{-n y} \leq n e^{-n ln(1+y)} (y-ln(1+y))=n(1+y)^{-n} (y-ln(1+y)) 
\end{equation}

for any real $y\geq 0$ and  integer $n\geq 1$

\medskip

The term $y-ln(1+y)$ appearing in the last term in the
inequality above is estimated as

\[
0\leq y-ln(1+y)\leq \frac{y^2}{2}
\]

for any real $y\geq 0$

\medskip

Going back to inequality (\ref{inq1}) we finaly have

\[ 
e^{-n ln(1+y)}-e^{-n y} \leq n(1+y)^{-n} \frac{y^2}{2} 
\]

We replace  $y=x/n$ and obtain

\[ 
\left(1+\frac{x}{n}\right)^{-n}-e^{-x} \leq n\left(1+\frac{x}{n}\right)^{-n} \frac{x^2}{2n^2}= x^2
\left(1+\frac{x}{n}\right)^{-n} \frac{1}{2n}  
\]

\medskip

But, as pointed out before, for any fixed $x$ the function $n\to
(1+x/n)^{-n} $ is decreasing, so we have that for $n\geq 2$

\[
\left(1+\frac{x}{n}\right)^{-n} \leq \left(1+\frac{x}{2}\right)^{-2} \leq \frac{4}{1+x^2} 
\] 

Therefore, for $n\geq 2$

\[ 
\left(1+\frac{x}{n}\right)^{-n}-e^{-x} \leq x^2
\left(1+\frac{x}{n}\right)^{-n} \frac{1}{2n}  \leq \frac{4x^2}{1+x^2}\frac{1}{2n} \leq \frac{2}{n} 
\]

For $n=1$ the left hand side of (\ref{transf_est}) becomes

\[ 
(1+x)^{-1}-e^{-x} \leq (1+x)^{-1}+e^{-x} \leq 2\]

for any $x\geq 0$, so the inequality (\ref{transf_est}) is valid for $n=1$ too.

\bibliographystyle{plain}
\bibliography{Biblio}

\def\ocirc#1{\ifmmode\setbox0=\hbox{$#1$}\dimen0=\ht0 \advance\dimen0
  by1pt\rlap{\hbox to\wd0{\hss\raise\dimen0
  \hbox{\hskip.2em$\scriptscriptstyle\circ$}\hss}}#1\else {\accent"17 #1}\fi}
  \def\polhk#1{\setbox0=\hbox{#1}{\ooalign{\hidewidth
  \lower1.5ex\hbox{`}\hidewidth\crcr\unhbox0}}}
\begin{thebibliography}{10}

\bibitem{BFR80}
J.~Bardina, J.H. Ferziger, and W.C. Reynolds.
\newblock Improved subgrid scale models for large eddy simulation.
\newblock {\em AIAA paper}, 80:1357, 1980.

\bibitem{BL11}
L.~Berselli and R.~Lewandowski.
\newblock Convergence of {A}pproximate {D}econvolution {M}odels to the mean
  {N}avier-{S}tokes equations.
\newblock {\em Annales de l'Institut Henri Poincare (C), Non Linear Analysis},
  29:171--198, 2012.

\bibitem{BIL2006}
L.~C. Berselli, T.~Iliescu, and W.~J. Layton.
\newblock {\em Mathematics of {L}arge {E}ddy {S}imulation of turbulent flows}.
\newblock Scientific Computation. Springer-Verlag, Berlin, 2006.

\bibitem{CLT2006}
Y.~Cao, E.~M. Lunasin, and E.~S. Titi.
\newblock Global well-posedness of the three-dimensional viscous and inviscid
  simplified {B}ardina turbulence models.
\newblock {\em Commun. Math. Sci.}, 4(4):823--848, 2006.

\bibitem{CSXF05}
F.K. Chow, R.L. Street, M.~Xue, and J.H. Ferziger.
\newblock Explicit filtering and reconstruction turbulence modeling for
  large-eddy simulation of neutral boundary layer flow.
\newblock {\em Journal of the Atmospheric Sciences}, 62(7):2058--2077, 2005.

\bibitem{DE2006}
A.~Dunca and Y.~Epshteyn.
\newblock On the {S}tolz-{A}dams deconvolution model for the large-eddy
  simulation of turbulent flows.
\newblock {\em SIAM J. Math. Anal.}, 37(6):1890--1902 (electronic), 2006.

\bibitem{MG00}
M.~Germano.
\newblock Fundamentals of {L}arge {E}ddy {S}imulation.
\newblock In R.~Peyret and E.~Krause, editors, {\em CISM Courses and Lectures
  395, Advanced Turbulent flow computations}. Springer, 2000.

\bibitem{TG19}
T.~H. Gronwall.
\newblock Note on the derivative with respect to a parameter of the solutions
  of a system of differential equations.
\newblock {\em Ann. of Math.}, 20(4):292Ð296, 1919.

\bibitem{GC03}
J.~{G}ullbrand and F.K. Chow.
\newblock The effect of numerical errors and turbulence models in large-eddy
  simulation of channel flow, with and without explicit filtering.
\newblock {\em Journal of Fluid Mechanics}, 495(323-341):323--341, 2003.

\bibitem{LL2006a}
W.~J. Layton and R.~Lewandowski.
\newblock On a well-posed turbulence model.
\newblock {\em Discrete Contin. Dyn. Syst. Ser. B}, 6(1):111--128 (electronic),
  2006.

\bibitem{LL2006b}
W.~J. Layton and R.~Lewandowski.
\newblock Residual stress of approximate deconvolution models of turbulence.
\newblock {\em Journ. Turbul.}, 7:1--21, 2006.

\bibitem{LR2012}
W.~J. Layton and L.~Rebholz.
\newblock {\em {A}pproximate {D}econvolution {M}odels of {T}urbulence
  {A}pproximate {D}econvolution {M}odels of {T}urbulence}.
\newblock Springer, Heidelberg, 2012.

\bibitem{KL92}
K.~L. {L}ele.
\newblock Compact finite different schemes with spectral-like resolution.
\newblock {\em Journ. Comp. Phys.}, 103:16--42, 1992.

\bibitem{RLbookNSE}
R.~Lewandowski.
\newblock {\em Approximations to the {N}avier-{S}tokes {E}quations}.
\newblock In preparation, release scheduled end 2012.

\bibitem{LM1968}
J.-L. Lions and E.~Magenes.
\newblock {\em Probl\`emes aux limites non homog\`enes et applications. {V}ol.
  1}.
\newblock Travaux et Recherches Math\'ematiques, No. 17. Dunod, Paris, 1968.

\bibitem{Sag2001}
P.~Sagaut.
\newblock {\em Large eddy simulation for incompressible flows}.
\newblock Scientific Computation. Springer-Verlag, Berlin, 2001.
\newblock An introduction, With an introduction by Marcel Lesieur, Translated
  from the 1998 French original by the author.

\bibitem{stanculescu}
I.~Stanculescu.
\newblock Existence theory of abstract approximate deconvolution models of
  turbulence.
\newblock {\em Annali dell'Universita di Ferrara}, 51(1):145--168, 2008.

\bibitem{SS03}
E.~{S}tein and R.~Shakarchi.
\newblock {\em Fourier Analysis : an introduction}.
\newblock Princeton University Press, 2003.

\bibitem{SAK2001}
S.~Stolz, N.~A. Adams, and L.~Kleiser.
\newblock An approximate deconvolution model for large-eddy simulation with
  application to incompressible wall-bounded flows.
\newblock {\em Phys. Fluids}, 13(4):997--1015, 2001.

\bibitem{Tem2001}
R.~Temam.
\newblock {\em Navier-{S}tokes {E}quations}.
\newblock AMS Chelsea Publishing, Providence, RI, 2001.
\newblock Theory and numerical analysis, Reprint of the 1984 edition.

\end{thebibliography}
\end{document}